# QOS CATEGORIES ACTIVENESS-AWARE ADAPTIVE EDCA ALGORITHM FOR DENSE IOT NETWORKS

Mohammed A. Salem[1], Ibrahim F. Tarrad[2], Mohamed I. Youssef [2] and Sherine M. Abd El-Kader[3]

[1]Department of Electrical Engineering, Higher Technological Institute, 10th of Ramadan City, Egypt
[2]Department of Electrical Engineering, Al-Azhar University, Cairo, Egypt
[3]Department of Computer Engineering, Electronics Research Institute, Giza, Egypt

## ABSTRACT

*IEEE 802.11 networks have a great role to play in supporting and deploying of the Internet of Things (IoT). The realization of IoT depends on the ability of the network to handle a massive number of stations and transmissions and to support Quality of Service (QoS). IEEE 802.11 networks enable the QoS by applying the Enhanced Distributed Channel Access (EDCA) with static parameters regardless of existing network capacity or which Access Category (AC) of QoS is already active. Our objective in this paper is to improve the efficiency of the uplink access in 802.11 networks; therefore we proposed an algorithm called QoS Categories Activeness-Aware Adaptive EDCA Algorithm (QCAAAE) which adapts Contention Window (CW) size, and Arbitration Inter-Frame Space Number (AIFSN) values depending on the number of associated Stations (STAs) and considering the presence of each AC. For different traffic scenarios, the simulation results confirm the outperformance of the proposed algorithm in terms of throughput (increased on average 23%) and retransmission attempts rate (decreased on average 47%) considering acceptable delay for sensitive delay services.*

## 1. INTRODUCTION

The Internet of Things (IoT) is a heterogeneous concept that combines many different technologies, application domains, equipment facilities, and different services, etc. In IoT, a huge number of sensors and devices are expected to be connected through Machine-to-Machine (M2M) communications which are anticipated to support many industries with different utilizations such as smart grids and cities, telemedicine applications, vehicular telematics, surveillance systems, and manufacturing [1]-[2]. According to Ericsson, the expected number of IoT Stations (STAs) is expected to be 23.3 billion worldwide in 2023. The digitalization of equipment, vehicles and different processes lead to an exponentially increasing in the number of connected STAs [3]. The density of the network could be about 1~10 devices/m².

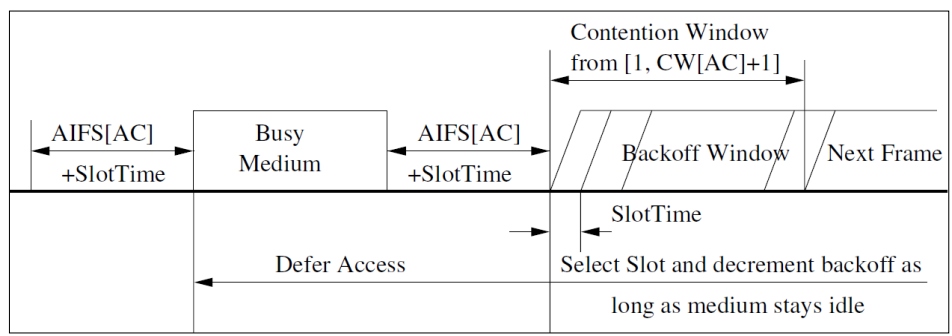

Figure 1. IEEE 802.11 EDCA Channel Access.

IEEE 802.11 networks have a great role to play in supporting and deploying of the IoT. The realization of IoT depends on the ability of the network to handle a massive number of stations and transmissions [4], and to support different QoS requirements for different types of IoT STAs and services. To support different QoS requirements, a prioritization mechanism called Enhanced Distributed Channel Access (EDCA) was introduced by IEEE 802.11 [5]. This mechanism provides four Access Categories (ACs): Voice (VO), Video (VI), Best Effort (BE), and Background (BK). The differentiation between ACs is provided by configuring different EDCA parameters [6], i.e. different Arbitration Inter-Frame Space Number (AIFSN) values and different Contention Window (CW) sizes for each AC; lower AIFSN and CW assigned for higher priority AC. As shown in Figure 1, once STA sensed the channel idle for a certain time slots equal to AIFSN, it generates a random back-off time within range from zero to the minimum contention window (CWmin), therefore starts to decrement the back-off time until it reaches to zero, at that time the STA starts to send its frame [7]. In case of the channel becomes busy during the decrement, the back-off time stops and wait till the channel becomes idle for AIFSN time slots again to continue the decrement. In case of unsuccessful transmission, the STA multiply the CW size by two; this multiplication continues until the CW size equals maximum contention window (CWmax). After any successful transmission, the STA resets the CW size to CWmin. Values of AIFSN and CW for each AC are illustrated in Table 1 [5]. Nevertheless, EDCA mechanism has many flaws due to static settings of AIFSN values and CW sizes regardless of existing network capacity condition or which AC of QoS is already existed. The increasing number of STAs led to a dramatically decreasing of network throughput, increasing collisions, higher delay due to increasing retransmissions [8].

Table 1. IEEE 802.11 EDCA Parameters.

| AC | AIFSN | CWmin | CWmax |
|----|-------|-------|-------|
| VO | 2 | 3 | 7 |
| VI | 2 | 7 | 15 |
| BE | 3 | 15 | 1023 |
| BK | 7 | 15 | 1023 |

The literature contains many works for overcoming the abovementioned flaws and to upgrade the performance of IEEE 802.11 networks. AIFSN values and CW size are very influential factors in determination of network throughput, mean average delay and packet retransmission attempts. Increasing of CW size led to lower collisions and retransmissions, but increases the network delay and vice versa; it's all about making a compromise between throughput and delay [9].

In traditional EDCA, IEEE sets the CW size to a static lower value for sensitive delay services, which mean a dramatic degradation of the performance in case of dense networks due to higher rates of collision and retransmission packets; this led to higher rates of delay [10]. In [11], the authors consider three possible traffic loads, their proposed scheme make an enhancement in terms of global throughput (overall network throughput) and retransmissions but degradation in terms of voice and video throughput. In [12], E. Coronado et al. introduce a dynamic AIFSN algorithm that adapts the AIFSN value depending on the network capacity to enhance the QoS. Many research attempts have been investigated to adapt the AIFSN such as [13], [14]. In [15], the authors provide an adaptive CW algorithm led to an improvement in the network performance, but not offered compatibility to legacy non-EDCA STAs. In [16], I. Syed et al. proposed an algorithm with providing an analytical model to adapt CW based on estimation of number of STAs for each AC considering the channel status, their proposed algorithm shows an enhancement in terms of global throughput and retransmission, but the calculations of CW show large CWmin value in relative to the number of stations. Of course with this large CW size; the terms of throughput and retransmission are expected to be improved, but the term of delay can be more improved. Many research efforts have been studied to adapt the CW size such as [17] – [19].

However, the previous proposals provided improvements, but without considering very high dense network condition and without considering the absence of any AC; which mean wasted resources that we can exploit it to improve the performance of the other ACs in terms of throughput, retransmission rate and delay for sensitive delay services.

For the abovementioned reasons, we proposed the QoS Categories Activeness-Aware Adaptive EDCA Algorithm (QCAAAE) algorithm for IEEE 802.11 Medium Access Control (MAC) to make EDCA parameters -especially CW size and AIFSN values- dynamically tuning, depending on the number of associated STAs in each AC, and considering the presence of each type of ACs. For different traffic scenarios, the simulation results confirm the outperformance of the QCAAAE algorithm in terms of throughput and retransmission rate considering acceptable delay for sensitive delay services. The remainder of this paper is structured as follows: Section 2 presents the QCAAAE algorithm in detail. Simulation parameters and performance evaluation of our work are detailed in Section 3. Finally, the conclusion is presented in Section 4.

## 2. PROPOSED ALGORITHM

AIFSN values and CW size are very influential factors in determination of network throughput, delay and packet retransmissions. Therefore, we proposed an adaptive algorithm to adjust AIFSN values and CW size considering the number of associated STAs in each AC, and the current presence of each AC to improve the efficiency of the uplink access in 802.11 networks (transmission from STAs to the Access Point (AP)). As shown in section III, the proposed algorithm enhances the performance of the network in terms of packet retransmissions, throughput considering acceptable delay for sensitive delay services. Our algorithm consists of 4 phases:

A) Determination of active ACs and number of associated STAs per each AC ($N_{AC}$)
B) Tuning of AIFSN values considering the absence of any AC.
C) Adaptation of CW size according to number of associated STAS per each AC.
D) Advertizing the new values of AIFSN and CW.

## 2.1. Determination of Active ACs and $N_{AC}$

Before STAs can transmit data through AP, they must be associated with AP to join the cell and take an Association Identifier (AID). During association process, every STA send an association request frame which contains QoS capability information as shown in Figure 2. Each STA sets AC flags (B0, B1, B2, and B3) to 1 to inform the AP of the required type(s) of QoS AC for sending and receiving data [5].

We calculated $N_{AC}$ for each category as follows: for every received association request frame, AP checks the flags of each required AC. In case of any AC flag found 1, the corresponding $N_{AC}$ increments. In contrast, for every disassociation request, the corresponding $N_{AC}$ decrements. Accordingly, any $N_{AC}$ has a value greater than 0 indicates that the corresponding AC is active. This mechanism is illustrated in Algorithm 1 and 2.

| B0 | B1 | B2 | B3 | B4 | B5  B6 | B7 |
|----|----|----|----|----|--------|----|
| AC_VO Flag | AC_VI Flag | AC_BK Flag | AC_BE Flag | Q-Ack | Max SP Length | More Data Ack |

Figure 2. QoS Capability Information Field for non-AP STA.

## 2.2. Tuning of AIFSN

As we mentioned before, the static allocation of AIFSN for different ACs is considered a waste of resources; especially in case of the absence of any AC. Therefore, lower priority ACs can improve its performance in terms of throughput, and delay by seizing opportunistically the resources of absent higher priority ACs. According to [20], most of the traffic in IoT networks is related to best effort category which has a priority lower than voice and video categories. Accordingly, in the case of absence of voice or video access categories, the BE access category has a real chance to improve its performance. Therefore, in our proposal, we take care of this issue and adapt AIFSN values with considering the absence ACs according to Table 2. For example, if voice and video ACs are inactive, then best effort AC will seize the minimum value of AIFSN which equal 2 to decrease its media access delay.

Table 2.  Proposed AIFSN values.

| AC Activity Status | | | AIFSN Values | | |
|--------------------|----------|----------|------|------|------|
| VO | VI | BE | VO | VI | BE |
| Inactive | Inactive | Active | - | - | 2 |
| Inactive | Active | Inactive | - | 2 | - |
| Inactive | Active | Active | - | 2 | 3 |
| Active | Inactive | Inactive | 2 | - | - |
| Active | Inactive | Active | 2 | - | 3 |
| Active | Active | Inactive | 2 | 3 | - |
| Active | Active | Active | 2 | 3 | 4 |

## 2.3. Adaptation of CW Size

In traditional EDCA, IEEE sets the CWmin and CWmax size to a static low value for each AC, which means a dramatic degradation of the performance in case of dense networks due to higher rates of collision and retransmission packets; this led to higher rates of delay. EDCA sets the initial CW size is to CWmin. After each unsuccessful transmission, the CW size is multiplied by 2 until it reaches to CWmax and saturated at this value. After each successful transmission, EDCA resets the CW size to CWmin.

In our proposed Algorithm and for each AC, we adapt the values of CWmin and CWmax according to the number of STAs in each AC. The new CWmin and CWmax values are given by equations (1) and (2).

$$CWmin[AC] = 2^{ceil(\log_2(\frac{N_{AC}}{2}))} - 1 \tag{1}$$

$$CWmax[AC] = \min(2^{ceil(\log_2(2N_{AC}))} - 1, PHY\_CWmax) \tag{2}$$

Where PHY_CWmax is the maximum size of CW restricted by the physical layer.

| Algorithm 1: Calculation of $N_{AC}$ | Algorithm 2: Update of AIFSN and CW |
|---|---|
| **Initialization:** | **Setting of AIFSN[AC]:** |
| $N_{VO} = 0$; $N_{VI} = 0$; $N_{BE}=0$; | 1:   if $N_{VO} > 0$, $N_{VI} > 0$ then |
| **Check re/assoc. request frame:** | 2:      AIFSN[VO] = 2 , AIFSN[VI] = 3, |
| 1:  if AC_VO_flag = 1 then |        AIFSN[BE] = 4 |
| 2:    $N_{VO} = N_{VO} +1$ | 3:  else if $N_{VO} > 0$, $N_{VI} = 0$ then |
| 3:  end if | 4:      AIFSN[VO] = 2 , AIFSN[BE] = 3 |
| 4:  if AC_VI_flag = 1 then | 5:  else if $N_{VO} = 0$, $N_{VI} > 0$ then |
| 5:    $N_{VI} = N_{VI} +1$ | 6:      AIFSN[VI] = 2 , AIFSN[BE] = 3 |
| 6:  end if | 7:  else |
| 7:  if AC_BE_flag = 1 then | 8:      AIFSN[BE] = 2 |
| 8:    $N_{BE} = N_{BE} + 1$ | 9:  end if |
| 9:  end if | **Settings of CW and Updating of EDCA:** |
| **Check disassoc. request frame:** | 1:  CWmin[VO]= 2^(ceil(log_2(N_{VO}/2))) - 1 |
| 1:  if AC_VO_flag = 1 then | 2:  CWmin[VI]= 2^(ceil(log_2(N_{VI}/2))) - 1 |
| 2:    $N_{VO} = N_{VO} - 1$ | 3:  CWmin[BE]= 2^(ceil(log_2(N_{BE}/2))) - 1 |
| 3:  end if | 4:  CWmax[VO]= min [2^(ceil(log_2(2*N_{VO}))) – 1, 1023] |
| 4:  if AC_VI_flag = 1 then | 5:  CWmax[VI]= min [2^(ceil(log_2(2*N_{VI}))) – 1, 1023] |
| 5:    $N_{VI} = N_{VI} -1$ | 6:  CWmax[BE]= min [2^(ceil(log_2(2*N_{BE}))) – 1, 1023] |
| 6:  end if | 7:  Update EDCA Parameters Field and send it in the |
| 7:  if AC_BE_flag = 1 then |       next scheduled Beacon Frame |
| 8:    $N_{BE} = N_{BE} - 1$ | |
| 9:  end if | |

## 2.4. Advertising the New AIFSN and CW Values

The AP broadcasts out periodically a beacon frame in the network, which contains all information about the network. Therefore, all STAs associated and connected to the network can be matched with the basic service set (BSS) parameters. Beacon frame contains a field for EDCA parameters setting.

In our proposed algorithm; after tuning of AIFSN value and calculation of CW size for each AC, the AP updates the EDCA parameters setting field and advertise all associated and connected STAs through the next scheduled beacon frame. Our proposed algorithm is illustrated in Algorithm 1, 2.

## 3. PERFORMANCE EVALUATION

In this section, we presented an evaluation of the performance of the proposed algorithm and compared it with the traditional EDCA. We evaluated our algorithm in Riverbed modeler (version 17.5) [21], and through simulation of a set of different 40 traffic scenarios. The simulation parameters are listed in Table 3. As we mentioned before, most of traffic in IoT networks is related to best effort category; so we created different scenarios as the following: 32, 64, 128, 256, and 512 BE stations, each scenario of these are repeated with 5, 15, 30 VO stations, 5, 15, 30 VI stations, and repeated with 15 VO & 15 VI stations; in all scenarios, each station always have load to send. The following metrics are used to figure out the performance of the proposed algorithm: Normalized throughput (the ratio between the throughput and the total loads submitted to the network), Mean Average Delay and Retransmission Attempts.

Table 3.  Simulation Parameters

| Parameter | Value | Parameter | Value |
|---|---|---|---|
| Physical Layer | IEEE 802.11n | Data Rate | 65 Mb/s |
| Spatial Streams | 1 | Guard Interval | 400 ns |
| Slot Time | 9 $\mu$s | VO Payload Size | 50 Bytes |
| AP Beacon Interval | 102.4 ms | VI Payload Size | 8738.13 Bytes |
| Physical CWmax | 1023 | BE Payload Size | 100 Bytes |

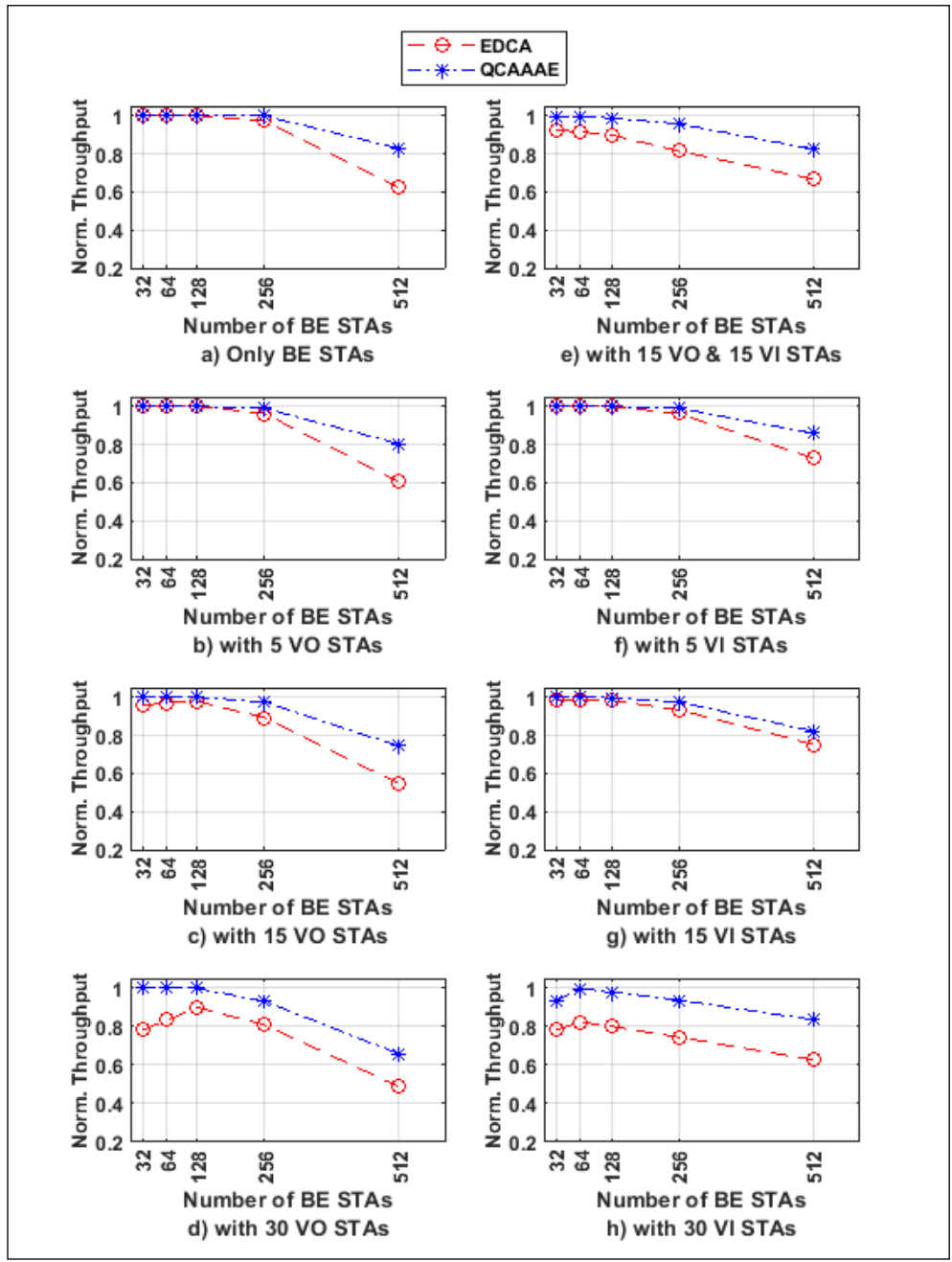

Figure 3. Global Normalized Throughput.

As shown in Figure 3, there is no doubt that the QCAAAE algorithm provides significantly more global normalized throughput than the traditional EDCA especially in high density scenarios. We also observed in Figure 4, that the global mean average delay in the proposed algorithm outperforms the traditional EDCA and higher in scenarios which contains many VI stations due to the higher bitrates of video stations; but on the other hand, the normalized throughput shows great enhancement on the QCAAAE algorithm.

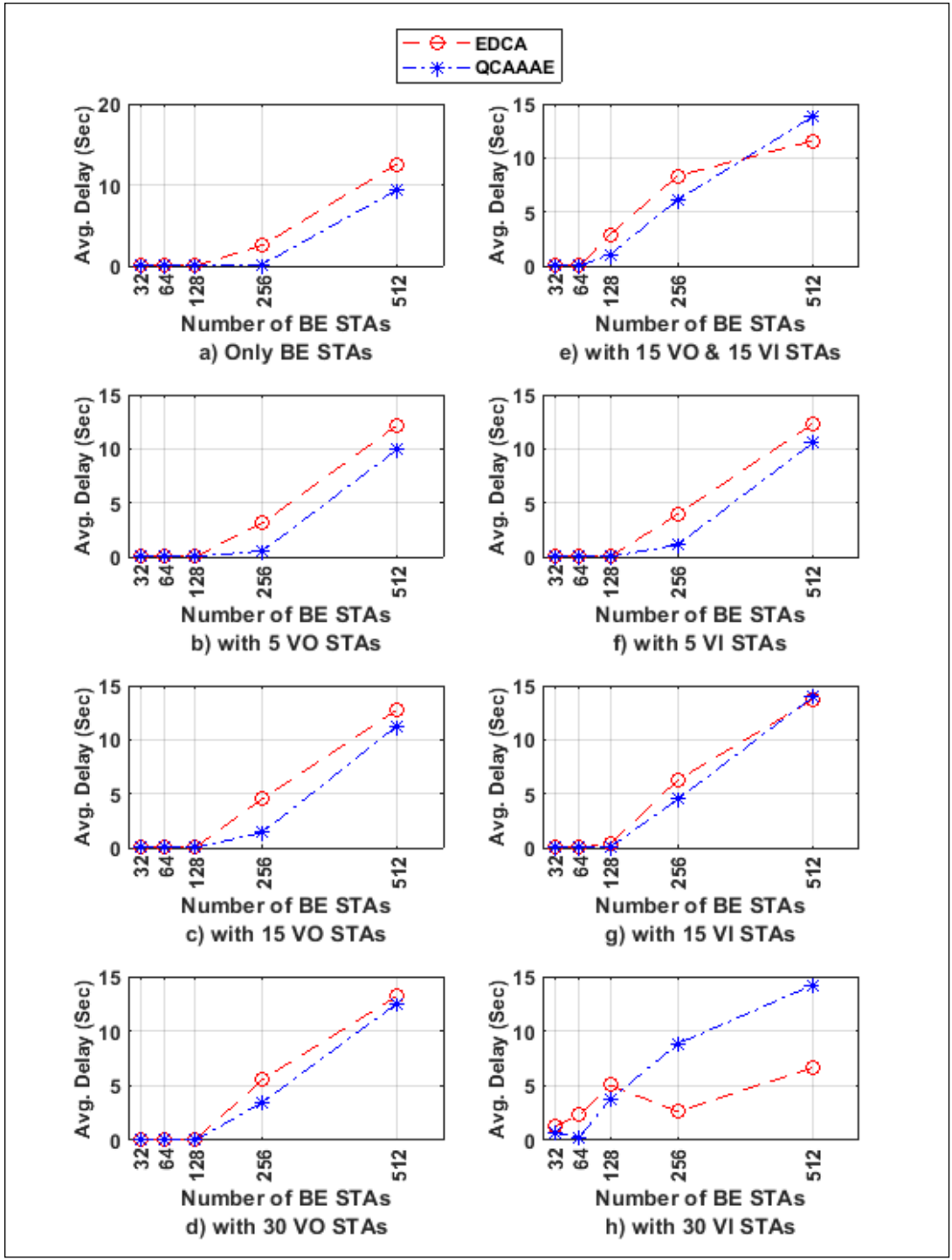

Figure 4. Global Mean Average Delay.

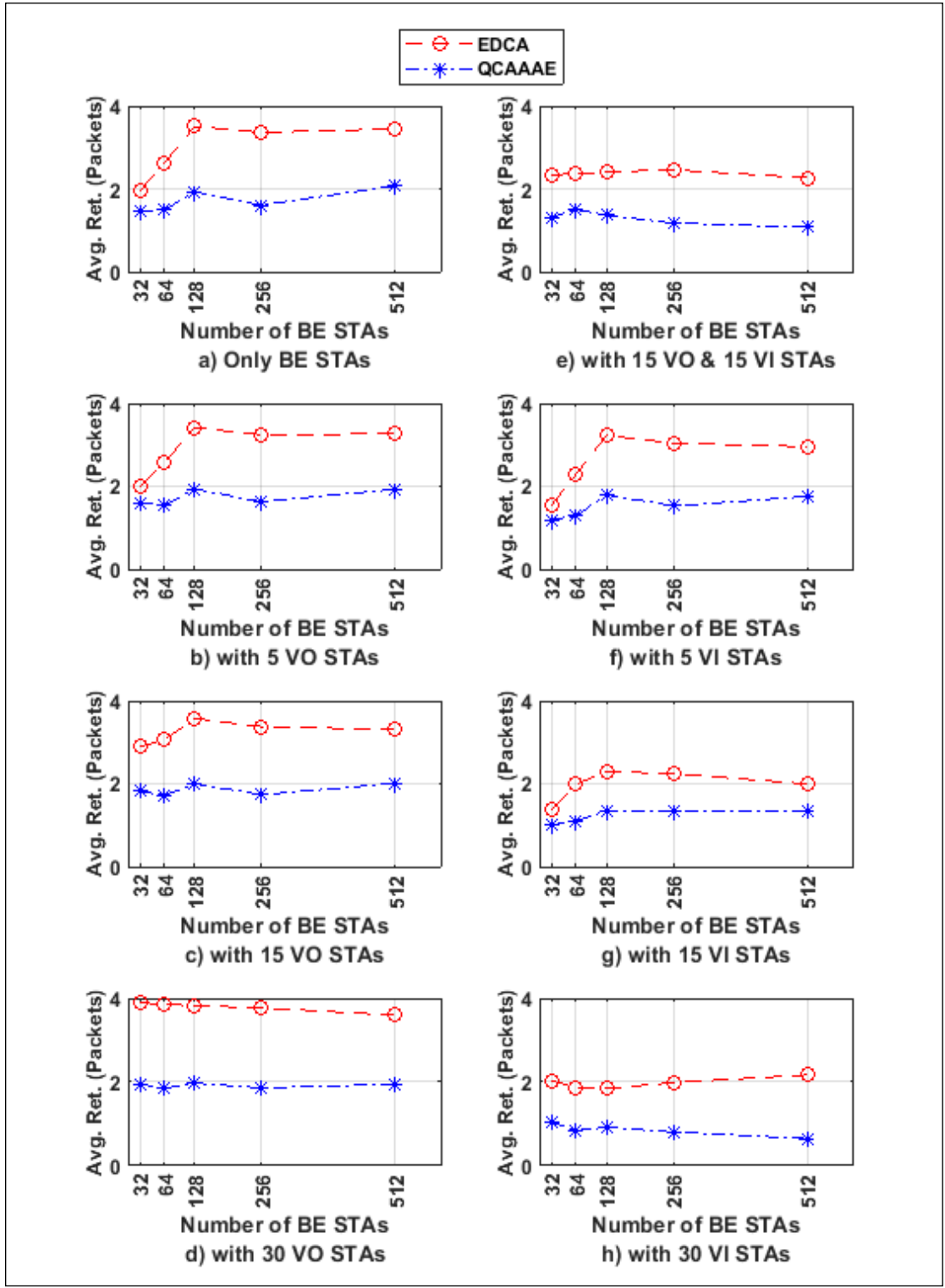

Figure 5. Global Retransmission Attempts.

We also observed in Figure 5, that the global retransmission attempts in the QCAAAE algorithm are always outperform the traditional EDCA in all scenarios because of the lower number of collision by the proposed algorithm. In Figure 6, it's shown that the proposed algorithm enhanced the normalized throughput of best effort AC greater than the traditional EDCA in all different traffic scenarios.

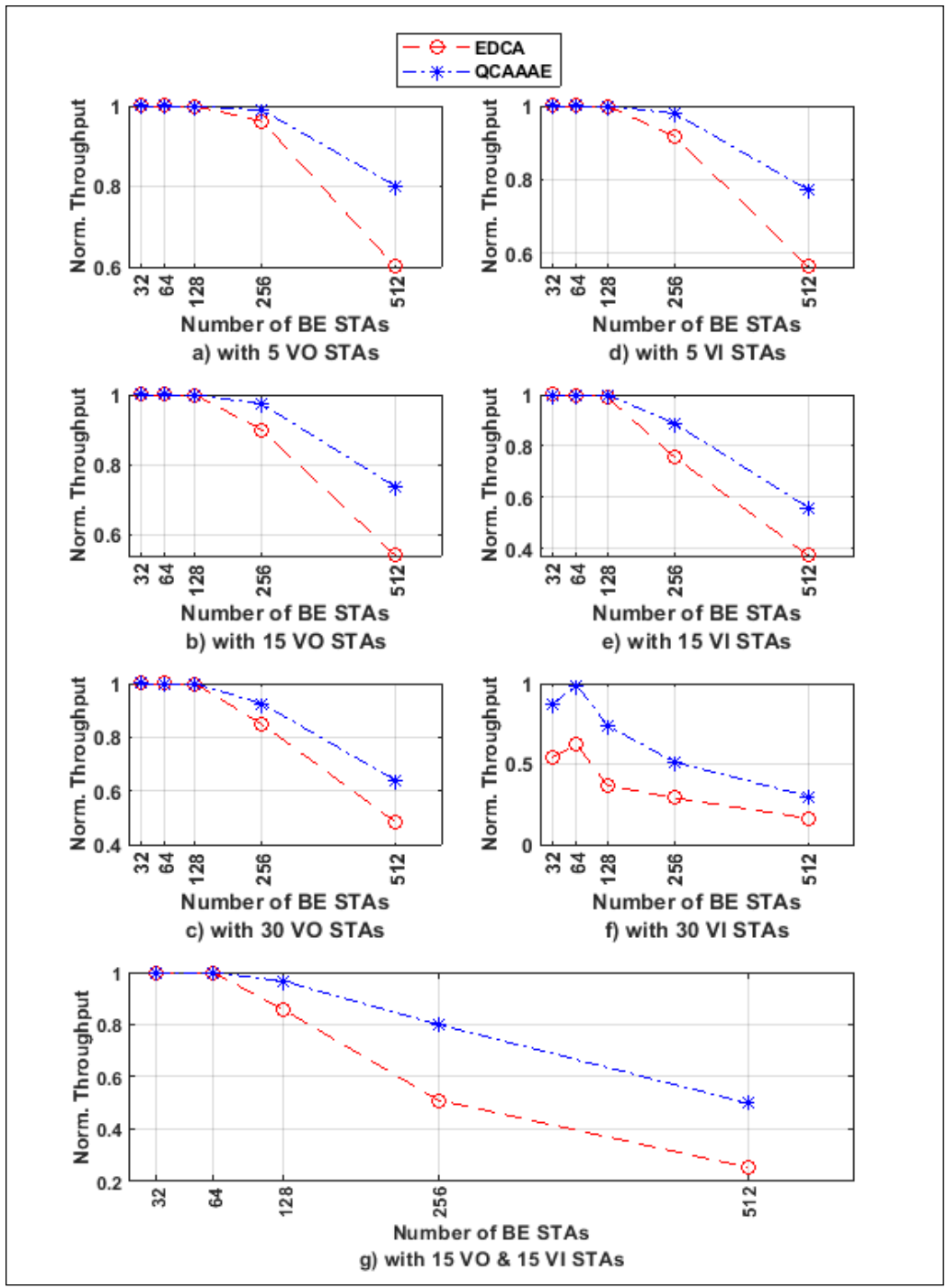

Figure 6. Best Effort AC Normalized Throughput.

According to [22], the preferred range of delay for voice applications is less than 150 ms, and the delay below 30 ms is not noticeable by the user. Figure 7 and Figure 8 shown the normalized throughput and mean average delay of voice AC, the mean delay of voice AC in the QCAAAE algorithm is higher than the traditional EDCA but still in the acceptable range while the normalized throughput of voice AC in the proposed algorithm shown magnificent improvement than EDCA. For Example, in the scenario of 30 voice stations & 512 best effort stations, the delay of voice AC rose from 3.8 ms to 7.5 ms; on the other hand, the normalized throughput improved significantly from 56.9% to 98.2%.

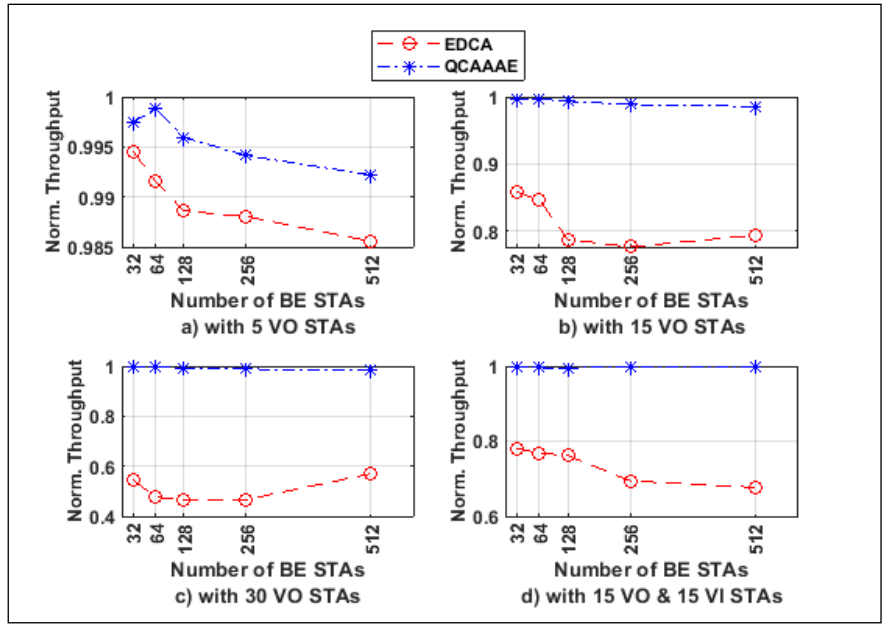

Figure 7. Voice AC Normalized Throughput.

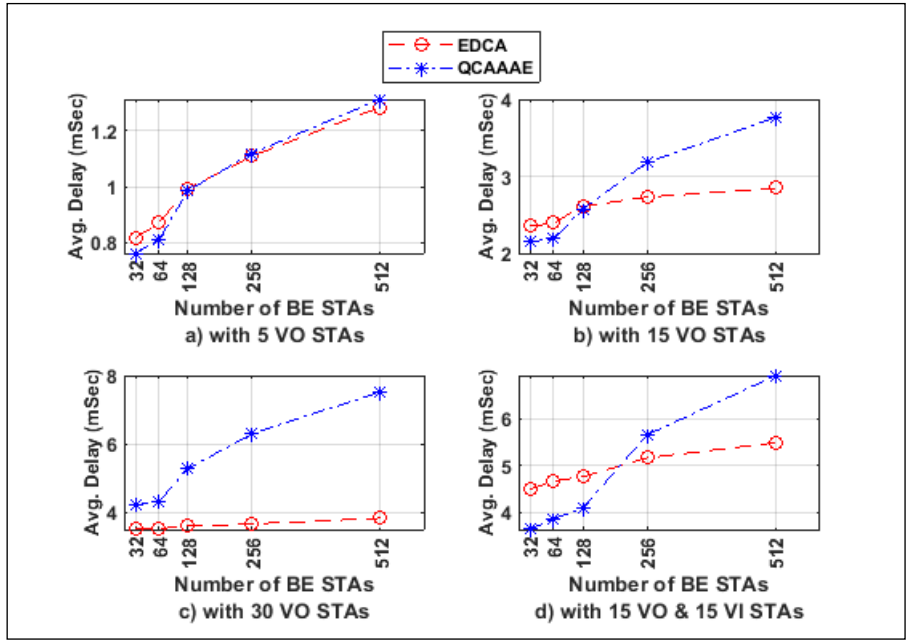

Figure 8. Voice AC Mean Average Delay.

According to [22], the preferred range of delay for video applications is less than 150 ms, Figure 9 and Figure 10 shown the normalized throughput and mean average delay for video AC, we observed that in some scenarios; the mean average delay in the proposed algorithm slightly rose within acceptable QoS requirements range; on the other hand, the normalized throughput of the proposed algorithm shown great improvements. For Example, in the scenario of 30 video stations with 512 best effort stations, the delay of video AC rose from 47.3 ms to 53.3 ms; on the other hand, the normalized throughput improved significantly from 75.1% to 97.9%.

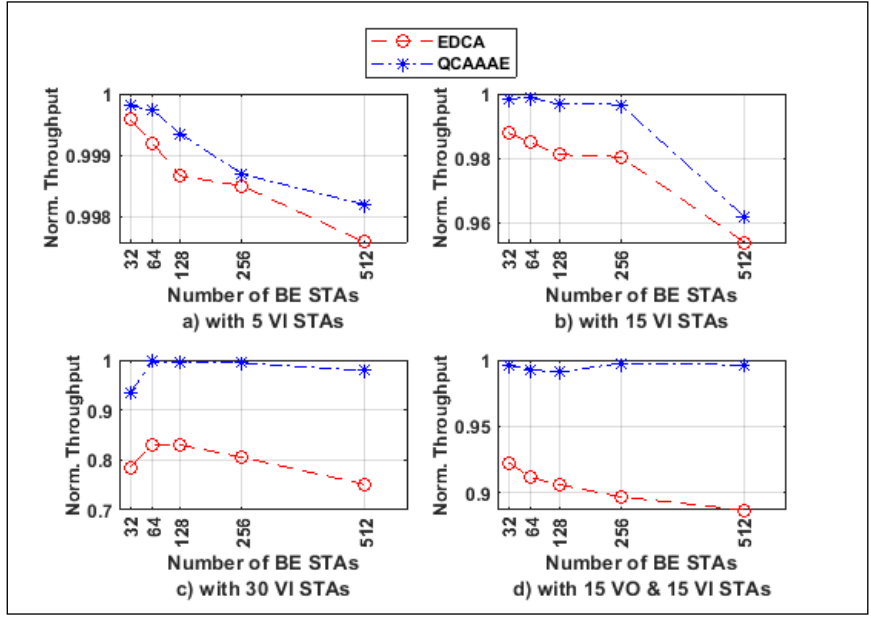

Figure 9. Video AC Normalized Throughput.

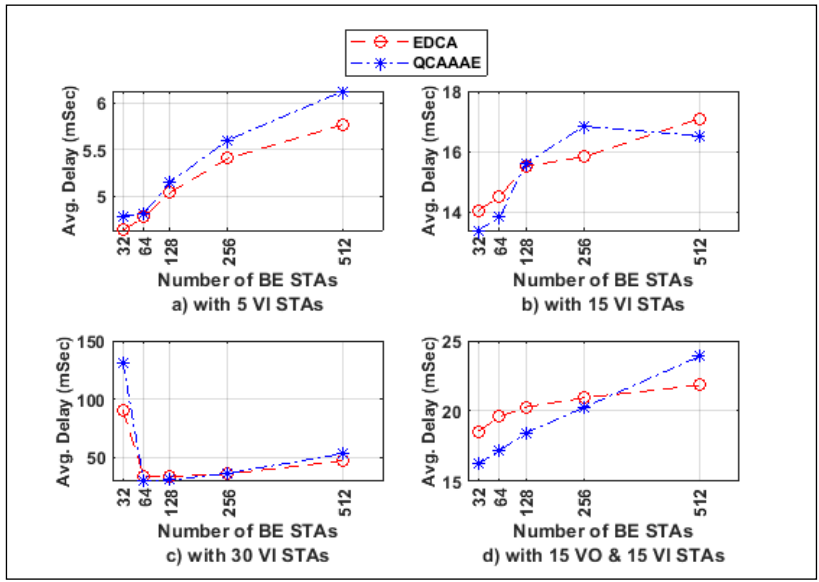

Figure 10. Video AC Mean Average Delay.

For better resolution reading, all previous simulation results are illustrated from Table 4 to Table 7.

Table 4. Global Normalized Throughput, Global Mean Average Delay, and Global Retransmission Attempts.

| Traffic Scenario | Global Normalized Throughput (%) | | Global Mean Average Delay (Sec) | | Global Retransmission Attempts (Packets) | |
|---|---|---|---|---|---|---|
| | EDCA | QCAAAE | EDCA | QCAAAE | EDCA | QCAAAE |
| **32 BE** | 100 | 100 | 0.004373 | 0.003554 | 1.9481 | 1.4823 |
| **64 BE** | 100 | 100 | 0.010194 | 0.006961 | 2.624 | 1.4855 |
| **128 BE** | 99.9987 | 100 | 0.02324 | 0.01602 | 3.5064 | 1.9327 |
| **256 BE** | 97.1780 | 99.7640 | 2.5717 | 0.1955 | 3.3609 | 1.6046 |
| **512 BE** | 62.2995 | 82.8831 | 12.564 | 9.383 | 3.4458 | 2.0738 |
| **32 BE, 5VO** | 99.9264 | 99.9652 | 0.004414 | 0.003864 | 2.0016 | 1.6072 |
| **64 BE, , 5VO** | 99.9389 | 99.9917 | 0.009957 | 0.007387 | 2.5805 | 1.5405 |
| **128 BE, 5VO** | 99.9557 | 99.9814 | 0.02334 | 0.01723 | 3.4126 | 1.9364 |
| **256 BE, 5VO** | 96.2243 | 98.9450 | 3.1337 | 0.5545 | 3.2467 | 1.6254 |
| **512 BE, 5VO** | 60.4104 | 80.3608 | 12.181 | 9.916 | 3.2647 | 1.9268 |
| **32 BE, 15VO** | 95.5170 | 99.9122 | 0.005674 | 0.004693 | 2.8902 | 1.8223 |
| **64 BE, 15VO** | 97.0947 | 99.9436 | 0.010833 | 0.008061 | 3.06 | 1.7235 |
| **128 BE, 15VO** | 97.7789 | 99.9256 | 0.02753 | 0.018422 | 3.5746 | 1.9984 |
| **256 BE, 15VO** | 89.2250 | 97.4479 | 4.5482 | 1.3966 | 3.377 | 1.7362 |
| **512 BE, 15VO** | 54.6561 | 74.3438 | 12.74 | 11.272 | 3.3077 | 2.008 |
| **32 BE, 30VO** | 78.1176 | 99.9192 | 0.00702 | 0.006232 | 3.8911 | 1.9515 |
| **64 BE, 30VO** | 83.3218 | 99.8939 | 0.012476 | 0.009432 | 3.8551 | 1.8507 |
| **128 BE, 30VO** | 89.8954 | 99.8231 | 0.04462 | 0.023212 | 3.8208 | 1.9732 |
| **256 BE, 30VO** | 80.8572 | 93.1582 | 5.5768 | 3.4711 | 3.7705 | 1.857 |
| **512 BE, 30VO** | 48.8139 | 65.8300 | 13.212 | 12.546 | 3.5972 | 1.9447 |
| **32 BE, 5VI** | 99.9616 | 99.9825 | 0.005029 | 0.004668 | 1.5439 | 1.1946 |
| **64 BE, , 5VI** | 99.9327 | 99.9776 | 0.010082 | 0.007717 | 2.2856 | 1.2897 |
| **128 BE, 5VI** | 99.9000 | 99.9519 | 0.0245 | 0.0179 | 3.2264 | 1.7911 |
| **256 BE, 5VI** | 96.2525 | 99.0359 | 3.9405 | 1.114 | 3.0408 | 1.5258 |
| **512 BE, 5VI** | 72.8739 | 85.8205 | 12.316 | 10.543 | 2.9559 | 1.755 |
| **32 BE, 15VI** | 98.8176 | 99.8510 | 0.013118 | 0.012073 | 1.3783 | 1.0096 |
| **64 BE, 15VI** | 98.5779 | 99.8827 | 0.017015 | 0.014095 | 1.9924 | 1.0916 |
| **128 BE, 15VI** | 98.2259 | 99.7167 | 0.3653 | 0.04552 | 2.298 | 1.3601 |
| **256 BE, 15VI** | 93.3207 | 97.3799 | 6.251 | 4.4876 | 2.2412 | 1.3429 |
| **512 BE, 15VI** | 75.1134 | 82.1623 | 13.721 | 13.949 | 1.9946 | 1.3341 |
| **32 BE, 30VI** | 78.1253 | 93.3843 | 1.268 | 0.6852 | 2.0291 | 1.0513 |
| **64 BE, 30VI** | 82.2663 | 99.6768 | 2.3 | 0.3025 | 1.8764 | 0.8419 |
| **128 BE, 30VI** | 80.0811 | 97.9876 | 5.067 | 3.7437 | 1.8454 | 0.9224 |
| **256 BE, 30VI** | 74.4371 | 93.7031 | 2.631 | 8.831 | 1.9856 | 0.8047 |
| **512 BE, 30VI** | 62.6733 | 83.4981 | 6.662 | 14.215 | 2.169 | 0.6478 |
| **32 BE, 15VO, 15 VI** | 92.2836 | 99.6146 | 0.016591 | 0.013143 | 2.3462 | 1.3157 |
| **64 BE, 15VO, 15 VI** | 91.4803 | 99.3157 | 0.03079 | 0.018855 | 2.3562 | 1.5041 |
| **128 BE, 15VO, 15 VI** | 89.8299 | 98.7993 | 2.908 | 1.0435 | 2.413 | 1.3619 |
| **256 BE, 15VO, 15 VI** | 81.3700 | 95.6391 | 8.338 | 6.1207 | 2.4556 | 1.1729 |
| **512 BE, 15VO, 15 VI** | 66.5864 | 82.4599 | 11.562 | 13.82 | 2.2666 | 1.0743 |

Table 5.  Best Effort Normalized Throughput.

| Traffic Scenario | Best Effort Normalized Throughput (%) | |
|---|---|---|
| | EDCA | QCAAAE |
| **32 BE** | 100 | 100 |
| **64 BE** | 100 | 100 |
| **128 BE** | 99.9987 | 100 |
| **256 BE** | 97.1780 | 99.7640 |
| **512 BE** | 62.2995 | 82.8831 |
| **32 BE, 5VO** | 100 | 100 |
| **64 BE, , 5VO** | 100 | 100 |
| **128 BE, 5VO** | 99.9982 | 99.9965 |
| **256 BE, 5VO** | 96.1742 | 98.9359 |
| **512 BE, 5VO** | 60.0397 | 80.1775 |
| **32 BE, 15VO** | 100 | 100 |
| **64 BE, 15VO** | 100 | 100 |
| **128 BE, 15VO** | 99.9895 | 99.9927 |
| **256 BE, 15VO** | 89.8973 | 97.3654 |
| **512 BE, 15VO** | 53.9329 | 73.6375 |
| **32 BE, 30VO** | 100 | 100 |
| **64 BE, 30VO** | 100 | 99.9968 |
| **128 BE, 30VO** | 99.9525 | 99.9761 |
| **256 BE, 30VO** | 84.8490 | 92.4968 |
| **512 BE, 30VO** | 48.3401 | 63.9446 |
| **32 BE, 5VI** | 100 | 100 |
| **64 BE, , 5VI** | 100 | 100 |
| **128 BE, 5VI** | 99.9860 | 99.9949 |
| **256 BE, 5VI** | 91.7813 | 98.0002 |
| **512 BE, 5VI** | 56.1669 | 77.1209 |
| **32 BE, 15VI** | 100 | 99.9975 |
| **64 BE, 15VI** | 99.9924 | 99.9885 |
| **128 BE, 15VI** | 99.0983 | 99.9563 |
| **256 BE, 15VI** | 75.7296 | 88.8628 |
| **512 BE, 15VI** | 37.3173 | 56.0246 |
| **32 BE, 30VI** | 54.2993 | 86.9301 |
| **64 BE, 30VI** | 62.0384 | 98.8532 |
| **128 BE, 30VI** | 36.7471 | 74.0623 |
| **256 BE, 30VI** | 28.9552 | 51.2297 |
| **512 BE, 30VI** | 16.2281 | 29.8907 |
| **32 BE, 15VO, 15 VI** | 99.9872 | 99.9910 |
| **64 BE, 15VO, 15 VI** | 99.9706 | 99.9835 |
| **128 BE, 15VO, 15 VI** | 85.7625 | 96.6277 |
| **256 BE, 15VO, 15 VI** | 51.1051 | 80.0190 |
| **512 BE, 15VO, 15 VI** | 25.3506 | 49.9526 |

Table 6.  Voice Normalized Throughput and Mean Average Delay.

| Traffic Scenario | Voice Normalized Throughput (%) | | Voice Mean Average Delay (ms) | |
|---|---|---|---|---|
| | EDCA | QCAAAE | EDCA | QCAAAE |
| **32 BE, 5VO** | 99.4531 | 99.7418 | 0.8162 | 0.7624 |
| **64 BE, , 5VO** | 99.1533 | 99.8844 | 0.8709 | 0.8116 |
| **128 BE, 5VO** | 98.8608 | 99.5942 | 0.993 | 0.9854 |
| **256 BE, 5VO** | 98.8033 | 99.4134 | 1.109 | 1.1188 |
| **512 BE, 5VO** | 98.5563 | 99.2203 | 1.2827 | 1.3116 |
| **32 BE, 15VO** | 85.9082 | 99.7241 | 2.3509 | 2.1505 |
| **64 BE, 15VO** | 84.6403 | 99.7020 | 2.3918 | 2.188 |
| **128 BE, 15VO** | 78.8264 | 99.3511 | 2.611 | 2.5773 |
| **256 BE, 15VO** | 77.6975 | 98.8617 | 2.729 | 3.179 |
| **512 BE, 15VO** | 79.4560 | 98.5665 | 2.8454 | 3.7644 |
| **32 BE, 30VO** | 54.6663 | 99.8325 | 3.4989 | 4.215 |
| **64 BE, 30VO** | 47.5739 | 99.6735 | 3.527 | 4.32 |
| **128 BE, 30VO** | 46.7827 | 99.1672 | 3.5958 | 5.277 |
| **256 BE, 30VO** | 46.6339 | 98.8282 | 3.664 | 6.296 |
| **512 BE, 30VO** | 56.9379 | 98.1580 | 3.8259 | 7.516 |
| **32 BE, 15VO, 15 VI** | 77.9914 | 99.7446 | 4.491 | 3.63 |
| **64 BE, 15VO, 15 VI** | 76.8547 | 99.5970 | 4.667 | 3.865 |
| **128 BE, 15VO, 15 VI** | 76.1537 | 99.5110 | 4.757 | 4.0867 |
| **256 BE, 15VO, 15 VI** | 69.3884 | 99.8852 | 5.165 | 5.657 |
| **512 BE, 15VO, 15 VI** | 67.6655 | 99.8266 | 5.488 | 6.93 |

Table 7.  Video Normalized Throughput and Mean Average Delay.

| Traffic Scenario | Video Normalized Throughput (%) | | Video Mean Average Delay (ms) | |
|---|---|---|---|---|
| | EDCA | QCAAAE | EDCA | QCAAAE |
| **32 BE, 5VI** | 99.9577 | 99.9808 | 4.631 | 4.78 |
| **64 BE, , 5VI** | 99.9192 | 99.9731 | 4.773 | 4.815 |
| **128 BE, 5VI** | 99.8654 | 99.9346 | 5.036 | 5.146 |
| **256 BE, 5VI** | 99.8500 | 99.8692 | 5.402 | 5.598 |
| **512 BE, 5VI** | 99.7577 | 99.8192 | 5.764 | 6.126 |
| **32 BE, 15VI** | 98.7780 | 99.8461 | 14.037 | 13.393 |
| **64 BE, 15VI** | 98.4831 | 99.8756 | 14.515 | 13.823 |
| **128 BE, 15VI** | 98.1089 | 99.6846 | 15.513 | 15.605 |
| **256 BE, 15VI** | 98.0385 | 99.6640 | 15.82 | 16.835 |
| **512 BE, 15VI** | 95.3864 | 96.1820 | 17.09 | 16.516 |
| **32 BE, 30VI** | 78.5246 | 93.4925 | 90.94 | 1.984 |
| **64 BE, 30VI** | 82.9444 | 99.7044 | 10.814 | 0.6335 |
| **128 BE, 30VI** | 82.9866 | 99.5917 | 20.038 | 9.741 |
| **256 BE, 30VI** | 80.5360 | 99.3986 | 6.803 | 19.259 |
| **512 BE, 30VI** | 75.1293 | 97.8750 | 17.111 | 29.497 |
| **32 BE, 15VO, 15 VI** | 92.2489 | 99.5999 | 18.554 | 16.237 |
| **64 BE, 15VO, 15 VI** | 91.1398 | 99.2665 | 19.604 | 17.171 |
| **128 BE, 15VO, 15 VI** | 90.5893 | 99.0793 | 20.283 | 18.44 |
| **256 BE, 15VO, 15 VI** | 89.6742 | 99.7618 | 20.941 | 20.249 |
| **512 BE, 15VO, 15 VI** | 88.6876 | 99.6245 | 21.843 | 23.89 |

# 4. CONCLUSION

Motivated by deploying of the QoS-empowered IoT Networks in this paper, we produced an algorithm called QoS Categories Activeness-Aware Adaptive EDCA algorithm (QCAAAE), which dynamically adapts AIFSN value and CW size according to the active QoS access categories and according to the number of associated stations per each access category to consider high dense networks and exploit wasted resources of the inactive access categories.

For different traffic scenarios and using Riverbed modeler, the obtained simulation results show that the QCAAAE algorithm improves the performance of the network more than the traditional EDCA for all QoS ACs in terms of normalized throughput (increased on average 23%), retransmission attempts (decreased on average 47%) and mean average delay with considering of acceptable delay for sensitive delay applications and services. In some traffic scenarios which contain a large number of stations, the mean delay of voice and video services slightly increased but still in the recommended acceptable range; on the other hand, the throughput of voice and video services greatly increased as shown in the simulation results.